\documentclass[a4paper]{article}

\usepackage{xcolor}

\usepackage{cite}

 \usepackage[dvips]{graphicx}
  \DeclareGraphicsExtensions{.eps}

\usepackage[cmex10]{amsmath}

\usepackage{array}

\usepackage{dblfloatfix}

\usepackage{url}

\begin{document}

\title{Epitaxial rhenium microwave resonators}

\author{
E. Dumur, B. Delsol, T. Wei\ss l, B. Kung, W. Guichard, C. Hoarau, \\
C. Naud, K. Hasselbach and O. Buisson \\
Universit\'e Grenoble Alpes and CNRS, Institut NEEL,\\
F-38000 Grenoble, FRANCE\\
Email: c\'ecile.naud@neel.cnrs.fr
\and
K. Ratter, B. Gilles \\
Universit\'e Grenoble Alpes and CNRS, SIMAP\\ 
F-38 000 Grenoble, FRANCE\\
Email: kitti.ratter@neel.cnrs.fr
}

\maketitle

\begin{abstract}
We have fabricated rhenium microwave resonators from epitaxial films. We have used thin films of different structural quality depending on their growth conditions.
The resonators were coupled to a microwave transmission line which allows the measurement of their resonance frequencies and internal quality factors.
From the resonance frequency at low temperature, the effective penetration depth and the London penetration depth of the rhenium film are extracted.
 
\end{abstract}


\section{Introduction}

This work has been performed in the frame of the realization of  superconducting quantum circuits based on Josephson junctions. These circuits can show quantized energy levels like an artificial atom. Formally, a quantum bit is a two-state quantum mechanical system. The first two levels of such artificial atoms can be considered as quantum bits. In order to manipulate the quantum bit, we typically couple such an artificial atom to photons of a microwave resonator.

Up to now, most of the superconducting circuits are realized using aluminium. These films are polycrystalline. High density of crystal defects induces scattering processes for the charge carriers
 and limits the coherence time of the quantum bits \cite{kline}. This is the reason why we are preparing our superconducting thin films by molecular beam epitaxy. The rhenium has been chosen because
 the lattice mismatch between rhenium and sapphire [0001] is small (about -0.47\% at room temperature in the in-plane direction $<01\overset{-}{1}0>$). The quality of the layer as well as the interface between sapphire and rhenium is very high \cite{haq}. Moreover, rhenium does not oxidise, leading to clean interfaces.

This article presents the realization and the low temperature characterization of quarter wave ($\lambda /4 $) resonators coupled to a transmission feedline made out of epitaxial rhenium films. The microwave characterization has allowed the determination of the London penetration depth for the first time to our knowledge. Our experiment is the first step toward the realization of quantum circuits using epitaxial rhenium. 

\section{Model of the system}
\begin{figure}[!t]
\centering
\includegraphics[width=2.2in]{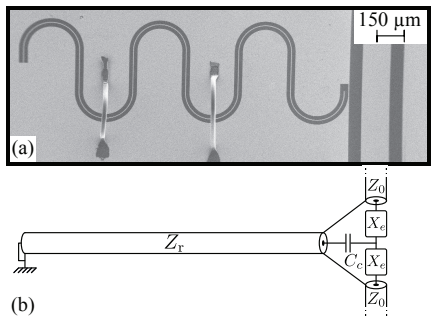}
\caption{(a) SEM picture of a $\lambda/4$ resonator coupled to a feedline.
The coupling is realised with an elbow coupling geometry.
(b) Equivalent electrical circuit.
The coupling is modelled as a single capacitor $C_c$ and the feedline as a transmission line of characteristic impedance $Z_0$.
The resonator is modelled as a transmission line of characteristic impedance $Z_r$.}
\label{fig1}
\end{figure}
\hyphenation{wave-guide}
The design of the sample is realized in the coplanar waveguide geometry.
A scanning electron microscopy (SEM) picture of a resonator coupled to a transmission feedline is shown in the figure 1 (a). The resonator of length $l=4,188~\mu m$ is formed of a center conductor
 of width $w=5~\mu m$ separated from the lateral ground planes by a gap of width $s=20~\mu m$. The center conductor is shorted to the ground at one side and opened at the other. A schematic view of the system is shown in the figure 1 (b).
 The resonator is modelled as a transmission line. 
 The capacitance per unit length $C_l$, the characteristic impedance $Z_r$, the magnetic inductance per unit length $L_l$ and the kinetic inductance per unit length $L_K$ of
 a resonator are the physical parameters describing its properties. They depend on the geometrical dimension of the resonator and the physical properties of the material
 (such as the effective permittivity of the substrate and the effective penetration depth of the superconducting layer)\cite{collin}. The resonance frequency $\omega_r$
 can be expressed through these parameters. In the case of a $\lambda /4 $ resonator, one has: 
 
\begin{equation}
    \frac{\omega_r}{2 \pi} = \frac{1}{4l} \sqrt{ \frac{1}{(L_l + L_K)C_l}}.
    \label{nu}
\end{equation}

\begin{figure}[!t]
\centering
\includegraphics[width=3.0in]{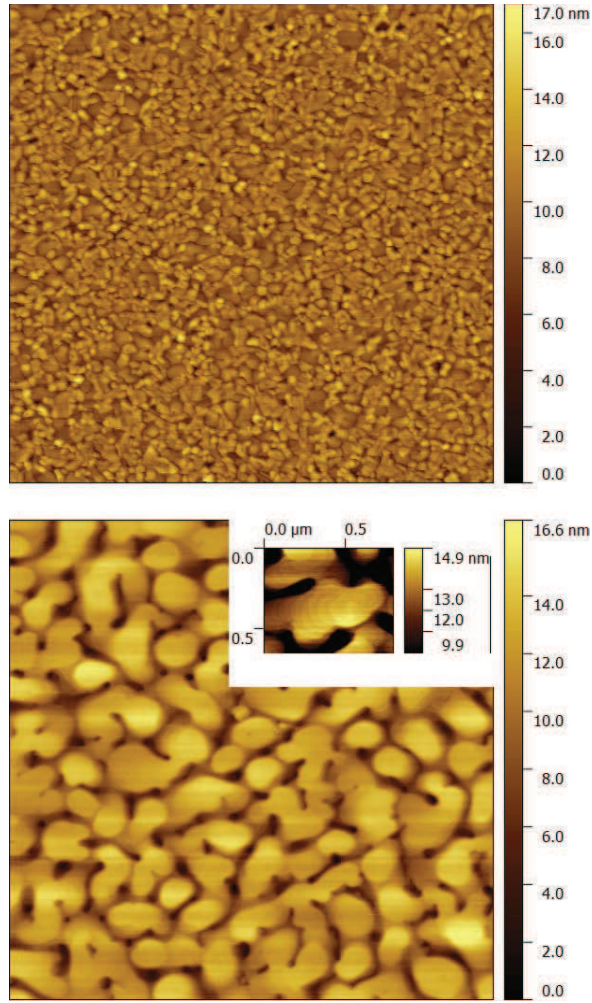}
\caption{3~$\mu $mx3~$\mu$m atomic force microscopy images of the surface of the two rhenium layers. Top: Sample A grown at 970~K. Bottom: Sample B grown at 1040~K. Inset : zoom on a spiral}
\label{fig_sim}
\end{figure}

Our resonators uses the "elbow" geometry to be coupled to the feedline \cite{mazin}. By this way, three resonators per chip are coupled to the feedline. 
The length of the elbow part is small compared to the resonator length and the microwave wavelength at the resonance frequency.
Only the part where the electric field is predominant is coupled to the feedline.
We can thus neglect the inductive coupling.
The capacitive coupling is modelled by a single capacitor $C_c$.
The resonance frequency of the resonator coupled to the feedline is denoted $\omega_0$. $\omega_0=\omega_r(1-Z_r C_c \omega_r/2)$ is slightly detuned compared to the bare resonance frequency $\omega_r$ because of the capacitive coupling. The relative drive frequency is $\Delta \omega_0 = \omega - \omega_0$ where $\omega$ is the applied angular frequency. 
The feedline has been made to have a characteristic impedance $Z_0=50 \Omega$.

The transmission through the feedline can be calculated at the vicinity of $\omega_0$  ($\frac{\Delta \omega_0}{\omega_0}\ll C_c \omega_0 Z_r/2$) \cite{dumur} and gives:
\begin{equation}
    S_{21} = \frac{1 + 2i Q_i\frac{\Delta \omega_0}{\omega_0}}{1 + \frac{Q_i}{Q_c} + 2i Q_i\frac{\Delta \omega_0}{\omega_0}}
    \label{eq:s21}
\end{equation}
where $Q_i$ is the internal quality factor and $Q_c$ is the external quality factor. The derivation of the calculation gives rise to: $Q_c=2/(C_c^2\omega_0 Z_r Z_0)$.
We have assumed a weak coupling: $Z_r C_c \omega_0/2\ll 1$ and $Z_0 C_c \omega_0/2\ll 1$.
The transmission of the signal amplitude is given by $S_{21}$ while $\left|S_{21}\right|^2$ corresponds to power transmitted.
$\left|S_{21}\right|^2$ as function of the frequency follows a Cauchy--Lorentz distribution shape:
\begin{equation}
    \left|S_{21}\right|^2 = 1 - \frac{1 - \left( \frac{Q_0}{Q_i} \right)^2}{1 +\left( 2 Q_0 \frac{\Delta \omega_0}{\omega_0}\right)^2} = 1 - \frac{I}{1 + \left(2 \frac{\Delta \omega_0}{\delta_{FWHM}} \right)^2}
    \label{eq:s21-lorentz}
\end{equation}
where $Q_0^{-1} = Q_i^{-1} + Q_{c}^{-1} $ is the total quality factor.
The depth of the resonance dip and the full width at half maximum are given by $I = 1 - ( Q_0/Q_i )^2$ and $\delta_{FWHM} =  \omega_0/Q_0$.

Measured resonance dips may have an asymmetric shape and not a Lorentzian shape as shown in the figure 3.
Such asymmetric shapes are attributed to an impedance mismatch between feedline and transmission line \cite{khalil, megrant}. To take into account this effect, we simply introduce 
two identical imaginary impedances $i X_e$ as indicated in the figure 1 (b).
The transmission formula at the vicinity of $\omega_0$ becomes:
\begin{equation}
S_{21} =  \frac{\widehat{Q}_{c,eff}}{Q_c}\frac{1 + 2 i Q_i\frac{\Delta \omega_0}{\omega_0}}{1 + \frac{Q_i}{\widehat{Q}_{c,eff}} + 2 i Q_i\frac{\Delta \omega_0}{\omega_0}}
\label{eq:s21-as}
\end{equation}
where $\widehat{Q}_{c,eff}=Q_c Z_0/(Z_0 +i X_e)=\left|\widehat{Q}_{c,eff}\right|e^{i\phi}$. We assume a weak coupling: $Z_r C_c \omega_0/2\ll 1$ and $|Z_0 + i X_e| C_c \omega_0/2\ll 1$. $\widehat{Q}_{c,eff}$ normalizes the transmission near the resonance through the factor $\widehat{Q}_{c,eff}/Q_c$. In addition $\widehat{Q}_{c,eff}$ plays a similar role than the $Q_c$ in Eq.\ref{eq:s21-lorentz} except that it is a complex number and determines
 the asymmetry as discussed in ref.\cite{khalil, megrant}.

\section{Realization of the $\lambda /4$ rhenium resonators}

Different samples of epitaxial rhenium have been fabricated. The growth conditions have been changed from sample to sample in order to find optimal parameters. The sapphire has been annealed at 1400~K in a 20\%-oxygen
 and 80\%-argon atmosphere to obtain atomically smooth surface of terrace and step like structure.
 Rhenium is grown under ultra-high vacuum in order to avoid contamination.
The typical deposition rate is 0.1~\AA .s$^{-1}$. An important parameter is the temperature of the substrate during the growth which ensures a good mobility of ad atoms on the substrate.
At low temperature (around 970~K), the surface is composed of grains as seen on the atomic force microscopy (AFM)
 picture in the figure 2-top. The reflection high energy electron diffraction (RHEED) pattern indicates a 2D layer and 3D grains.
 The X-ray diffraction shows one large rhenium peak corresponding to the epitaxial orientation and an other smaller peak corresponding to a secondary orientation confirming
 that part of the surface is not epitaxial. The rhenium layer $A$ was 25~nm thick.
It exhibits a resistivity at room temperature of 21.2~$\mu \Omega$.cm and a residual resistance ratio (RRR) of 4.45.
The intrinsic coherence length and the mean free path have been extracted  by transport measurements and BCS theory \cite{tulina}.
We obtain  $\xi_0$ = 160~nm and $l_e$ = 45~nm, respectively. The transition between superconductor and normal metal is unusual with a kind of two sub-transitions, the first one at about 1.85 K and the second one at 2.125~K.

For a growth at higher temperature (1 040 K), we can observe atomic steps and spirals on the surface using AFM (see on the figure 2-lower and inset). These spirals are thought to arise from screw dislocations. 
The RHEED pattern shows rods and even Kikuchi lines corresponding to a single crystalline surface,  confirmed by X-ray diffraction. The layer $B$ was 100~nm thick.
It exhibits a resistivity at room temperature of 21.6~$\mu \Omega$.cm and a RRR of 40.
The intrinsic coherence length and the mean free path are $\xi_0$ = 146~nm and $l_e$ = 400~nm, respectively. The critical temperature is 1.88~K.

In order to fabricate the resonators, a resist mask was patterned onto the films by laser beam lithography. Then, a SF6 plasma etches the unprotected film. The final step consists in the removal of the remaining resist through acetone.

\begin{figure}[h!]
\centering
\includegraphics[width=3.0in]{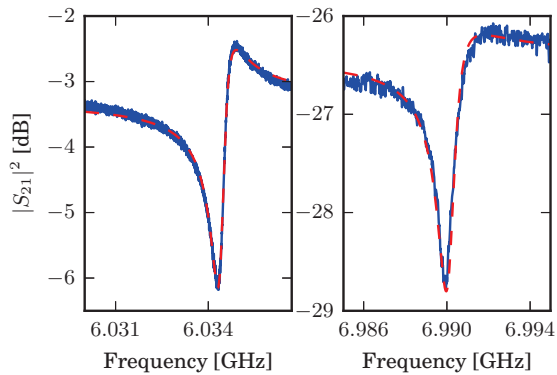}
\caption{Left: resonance corresponding to one resonator of the sample A. Right: resonance corresponding to one resonator of the sample B. Measurement was performed with a commercial VNA R\&S ZVL13 with a - 30~dBm output power. Values of $|S_{21}|^2$ are plotted with arbitrary offset in the vertical scale. The base temperature was T = 50~mK. Fit parameters for sample A gives $\frac{\omega_0}{2 \pi}$ = 6.0346~GHz, $\phi$=0,96~rad, $|\widehat{Q}_{c,eff}|$=29,000 and $Q_i$=14,000. Fit parameters for sample B gives $\frac{\omega_0}{2 \pi}$ = 6.99~GHz and $\phi$=0.59~rad, $|\widehat{Q}_{c,eff}|$=23,000 and $Q_i$=8,000.} 
\label{fig:multiplexing}
\end{figure}

\section{Experimental results}

Several microwave resonators based on the layers A and B have been characterized.
The figure 3 shows the microwave transmission measurements of these. They were performed in high power limit but still inside the linear response of the resonator.
We observe dips, each of them corresponding to one microwave resonator. The measured transmission coefficient has been fitted by the expression given in equation \ref{eq:s21-as}. 
The data are displayed in blue points and the model is plotted in dashed red.
Our model contains four independent parameters, namely the internal quality factor $Q_i$, the resonance frequency $\frac{\omega_0}{2 \pi}$, and the complex coupling factor $\widehat{Q}_{c,eff}$. These extracted parameters are listed in the caption of the figure \ref{fig:multiplexing}

\subsection{Extraction of the superconducting penetration depth}

From the resonance at the lowest temperature, the kinetic inductance and consequently the penetration depth can be extracted using Eq. \ref{nu}. Indeed, the geometric inductance and capacitance per unit length for a coplanar waveguide geometry are given by Ref.\cite{Gupta}:
\begin{equation}
    \label{LlCl}
    L_l = \frac{\mu_0}{4} \frac{K'(k)}{K(k)} \: \textrm{and} \: C_l=2\epsilon_0 (\epsilon_r+1) \frac{K(k)}{K'(k)} 
\end{equation}
with $k= w/(w+2s)$ and $K(k)$ is the complete elliptic integral of the first kind.
From Eq. \ref{nu} and \ref{LlCl}, we extract $L_k(0)=410~nH/m$.
To validate our method, we performed an ADS simulation of our circuit. The numerical values of  $L_l$ and $C_l$ are in good agreement with the extracted ones with a discrepancy less than 2\%.
From the quality factor obtained by the simulation considering zero losses, the coupling capacitance can be estimated: $C_c=2fF$. This coupling capacitance introduces an error in the extraction of $L_k(0)$ due to the frequency shift of the resonance frequency. Thus, the total error on the $L_k(0)$ extraction is estimated to be smaller than 5\%.
The kinetic inductance of a thin film ( $t \ll \lambda_{eff}$) can be written as \cite{barends2009photon}:
\begin{equation}
L_k(T) = \mu_0 g \frac{\lambda_{eff}^2 \left(T\right)}{t} 
\label{eq:lkk}
\end{equation}
where $t$ is the thickness and $\lambda_\text{eff}$ the effective penetration depth.
The parameter $g$, which allows the calculation of the kinetic inductance for a coplanar waveguide geometry, is defined as follow \cite{last, last2}:
\begin{multline}
g = \frac{1}{32 K^2\left(k\right)}\frac{\left( w + 2s\right)^2}{s\left(w+s\right)} \Big[ \frac{2}{w}\ln \left( \frac{w}{\delta}\frac{s}{w + s} \right) \\
 + \frac{2}{w + 2s}\ln\left(\frac{w+2s}{\delta} \frac{s}{w+s} \right)\Big],
 \label{oup}
\end{multline}
with the width $\delta = t/(4\pi e ^\pi)$. 
From Eq. \ref{eq:lkk} and \ref{oup}, the effective penetration depth is deduced: $\lambda_{eff} (0) = 180~nm$.
Since the rhenium film A is poorly epitaxial, the superconductivity is in the so-called \textit{dirty} limit.
The London penetration depth is related to the effective penetration depth as \cite{van1981}:
\begin{equation}
\lambda_\text{L}(0) = \lambda_\text{eff}(0)\sqrt{\frac{\xi_{eff}}{\xi_0}}
\label{eq:leff}
\end{equation}
with $\xi_{eff}^{-1} = \xi_0^{-1} + \ell_{e}^{-1}$.
Then we obtain: $\lambda_L(0) = 85 \pm 4~nm$. The penetration depth extraction has been obtained within a 5\% error using sample A. This is due to the thin thickness of the layer
and its relative large $\lambda_{eff} (0)$ which produces a kinetic inductance of the order of the geometric inductance.
This method could not be applied on layer B resonator whose kinetic inductance is estimated to be 10 times smaller and leading to a too large error.

\begin{figure}[t!]
\centering
\includegraphics[width=3.2in]{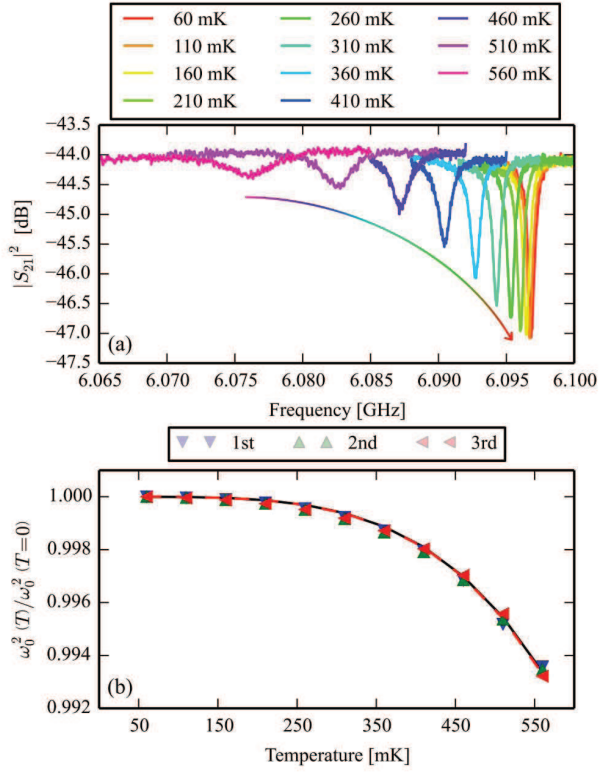}
\caption{Top: Temperature dependence of a resonance.
Bottom: Relative resonance frequency squared $\omega_0^2 (T)/\omega_0^2 (0)$ versus temperature for the three resonators (blue, red and green points).
The dashed black and red curves are calculated from Eq. \ref{eq:relative-frequency} with $\alpha_k = 0.8$, $T_c$ = 1.9~K and $\alpha_k = 0.4$, $T_c$ = 1.6~K, respectively.
We remark that in this range of temperature, the two theoretical curves are indistinguishable.}
\label{fig:re-temperature}
\end{figure}

Hereafter we discuss an alternative method to extract the kinetic inductance. This is based on the temperature dependence of the resonance.
In order to observe the effect of the temperature, we perform a series of resonance line shape measurements for the three resonators of sample A from 60~mK to 560~mK.
The line shapes of the three resonances exhibit the same behaviour with a shift to smaller frequency and a reduction of the dip depth when the temperature increases.
The relative frequencies squared $\omega_0^2 (T)/\omega_0^2 (0)$ versus temperature is shown in the figure 4 for the three resonators.
From equation \ref{nu}, we calculate the relative resonance frequency squared as:
\begin{equation}
\frac{\omega_0^2 (T)}{\omega_0^2 (0)} = \frac{L_\ell + L_\text{k} \left(0\right)}{L_\ell + L_\text{k}\left(T\right)}.
\end{equation}
Assuming Gorter Casimir law, the temperature dependence  of the kinetic inductance is given by:
\begin{equation}
L_k \left(T\right) = \frac{L_k\left(0\right)}{1 - \left(\frac{T}{T_c}\right)^4} .
\end{equation}
Leading to the relative resonance frequency temperature dependence:
\begin{equation}
\frac{\omega_0^2 (T)}{\omega_0^2 (0)} = \frac{1 - \left(\frac{T}{T_c}\right)^4}{1 - \left(\frac{T}{T_c}\right)^4 \left(1 - \alpha_k \right)} 
\label{eq:relative-frequency}
\end{equation}
with $\alpha_k = L_\text{k} \left(0\right)/(L_\ell +L_\text{k} \left(0\right))$ the fraction of kinetic inductance.
Data shown in the figure \ref{fig:re-temperature} are very well fitted by this equation with $T_c=1.6~K$ and $\alpha_k=0.4$.
However, due to the small dependence of the resonance frequency in this experiment, the uncertainty on the fit parameters are very large.
Indeed a broad range of parameters $(T_c,\alpha_k)$ can fit the data.
This is explained by the low temperature range of the experiment, where frequency changes by only 1\%. Such uncertainty can be strongly reduced by measuring $T_c$.
However, we were not able to extract $T_c$ due to the double superconducting transition.
In the future this method may be strongly improved in precision by measuring $T_c$ by transport measurement and by increasing the temperature range of the measurements.

\subsection{Low internal quality factor}

The extracted internal quality factors for sample A and B are about $14,000$ and $8,000$ respectively.
They present very close values and seems independent of the crystallographic quality suggesting that the limiting factor is not the density of defects. Moreover our measured Q factors are one order of magnitude lower than state of the art Q factors on epitaxial rhenium \cite{wang, sage}. In these reported measurements, the sample was placed inside a cylindrical cavity made of copper with 15~GHz cut-frequency. Metglas ribbon wraps the sample holder but no blackbody absorber was used in the present  measurements. We are currently improving the radiation shielding as well as the magnetic screening to reduce losses due to vortex motion and quasiparticles in order to improve the quality factor measurements \cite{barends2009photon}.

\section{Conclusion}

We have fabricated microwave resonators from an epitaxial layer of rhenium onto sapphire substrate. We succeed to realize state of the art epitaxial rhenium films with a RRR of 40. This was obtained when the deposition temperature is about 1,040~K. By measuring the resonance frequency of a lower quality layer deposited at 970~K, the kinetic inductance was large enough to be extracted with precision. The effective penetration depth and the London penetration depth have been deduced. Intrinsic quality factor on epitaxial rhenium microwave resonators remains an open issue.

\section*{Acknowledgment}

The authors would like to thank Bruno Fernandez for his help during the fabrication. The research has been supported by ANR-NSFC QUEXSUPERC, Grenoble Nanoscience Fondation and the 'laboratoire d'Excellence' LANEF.

\end{document}